# A PHYSICS OF BOUNDED METRICS SPACES

Pierre PERETTO
Département de Recherche Fondamentale sur la Matière Condensée
SP2M/PSC,
CEA Grenoble
17 rue des Martyrs, 38054 Grenoble Cedex 09, France
Mail:peretto@drfmc.ceng.cea.fr

**Abstract:** We consider the possibility of obtaining emergent properties of physical spaces endowed with structures analogous to that of collective models put forward by classical statistical physics. We show that, assuming that a so called « metric scale » does exist, one can indeed recover a number of properties of physical spaces such as the Minkowski metric, the relativistic quantum dynamics and the electroweak model.

### I)-Introduction

A mathematical object is fully determined by its information content. As information may be expressed by a number of bits, one can say that a mathematical object is fully determined by a set of bits. The purpose of theoretical physics, on the other hand, being describing physical systems as mathematical objects, it is to be admitted that, at the most foundamental level of description, the definition of a physical system can be reduced to a countable (eventually an infinite countable) set of bits. We call « *cells* » these elementary units of information. The cells are identified by an index (an address) $k = 1,2,\cdots$, and their internal states is taken as $\sigma_k \in \{-1,+1\}$. This way of considering physics fits the usual reductionnist approach which is to explain complex phenomena that are observed at a certain scale by the interplay of simpler phenomena occuring at a lower scale. Apparently high energy physics, which appeals to more and more complex spaces (10 dimensional spaces for fermions and 26 dimensional spaces for bosons in string theories for examples) goes the other way around. The Planck scale $l_{Planck} = \left(\dfrac{G_g \hbar}{c^3}\right)^{1/2} = 1.62 \times 10^{-35} m$ ($G_g$ is the universal gravitation constant, $c$ is the speed of light and $\hbar$ is the Planck constant) is the smallest scale which can be given a physical meaning. In high energy physics this is the scale where the supernumerary dimensions fold up. Here we rather consider that this is the scale of cells, that is to say the scale where information itself looses its meaning. Obviously it is to be proved that physics as it is known nowadays, may be reconstructed upon such bases. A few steps of this program are considered in this article.

### II)-Sites and cells

Information on cells, as defined above, is reduced to an address that is to say an identification index $k$ and to a (bivalent) internal state $\sigma_k$. So far cells are given no geometrical properties. Space-time for example must be an emergent property of the model not a property of physical systems that would be postulated *a priori*.

The cells being devoid of geometrical properties, there must exist a scale, which we call the metric scale, where the usual Minkowski properties of space-time are recovered. Just as matter looses its chemical properties under the atomic scale ($1eV \cong 10^{-10} m$), space looses its metric properties under the metric scale. The smallest part of space-time where the usual properties of (Minkowski) physical space can still be recognized we call a « *site* », a sort of space-time atom.

To summarize, the structure of physical spaces, as it is defined in the model we put forward in this article, is made of a countable set of cells and cells are divided up among sites. We call $i = 1,2,\cdots,N$ the index of a site, $\lambda = 1,2,\cdots,q$ the index of cells in a given site. $\sigma_{i,\lambda} \in \{-1,+1\}$ is the state of cell $(i\lambda)$. There is no geometry inside a site and, therefore any cell of a site is close to any other cell of the same site, that is to say all distances between the cells of a given site vanish. A site is similar to a geometrical point so to speak and the problem is that of understanding how the usual geometrical properties of physical spaces develop from such an organization.

Naturally it would be interesting to know more about the metric scale, if such a scale really does exist. It is certainly higher than the scale of some *100 GeV* which is now avalaible in the most powerful machines. Many high energy physicists, on the other hand, think that the coherence of present field theories is lost above a scale of, say, *1* to *10 TeV* ($1 TeV \cong 10^{-22} m$), which is the scale where the Higgs vacuum becomes unstable. These energies are now out of reach although those attainable in the LHC (Large Hadron Collider), a machine which is to be built at CERN, would be sensitively closer.

Astrophysics is the only domain where so large energies could eventually be observed. The $\gamma$ ray spectra of quasars display anomalies in the estimated metric scale range of energy indeed. For example in the high energy gamma ray emission spectrum of the active galactic nucleus (AGN) Markarian 421, the flux of *1 TeV* photons is only $10^{-5}$ of that of *1 GeV* photons (M.S. Schubnell et al.[1]). This is a general phenomenon which several theories strive to account for. According to the most accepted one, this anomaly is not to be related to the emission mechanism but, rather, to a degradation process implying interactions between the high energy photons and the infra-red and optical photons that are present in galactic and intergalactic spaces. This explanation implies that the spectra must strongly depend upon the distances of quasars and therefore it is not really satisfactory. This obviously does not mean that a satisfactory astrophysical explanation could not be found in the future but, for the moment, this leaves room for the metric scale hypothesis to be an acceptable one.

### III)-Intra- and inter-site physics

It is generally assumed that Nature obeys to some general principles and that not all states $I = \{\sigma_{i\lambda}\}$ of a physical system are equally probable. Some are preferred which amounts to saying that one can build a functional of states, namely a mapping of states in the set of real numbers, such as the preferred states correspond to minimal values of the functional. Those minima determine the dynamics of the physical system. Since time has been given no meaning so far, it is not legitimate to call this functional an Hamiltonian (or energy) although we shall do that later for the sake of convenience. The most simple, non trivial, functionnal is a quadratic form of states that is:

$$H(I) = -\sum_{kl} J_{kl} \sigma_k(I) \sigma_l(I)$$

The model, so defined, is similar to the Ising model of statistical mechanics (more precisely to a spin glass version of the Ising model). In terms of cells and sites the functionnal $H$ is rewritten as:

$$H = -\sum_{ij,\lambda\mu} J_{ij,\lambda\mu} \sigma_{i,\lambda} \sigma_{j,\mu}$$

where $J_{ij,\lambda\mu}$ couples cell $\lambda$ of site $i$ to cell $\mu$ of site $j$. As all mutual distances of cells belonging to a given site vanish, all cells of a site play the same role as regards the inter-site interactions and, therefore, the interactions factorize according to:

$$J_{ij,\lambda\mu} = K_{ij} M_{\lambda\mu}$$

or

$$J = K \otimes M$$

where $K$ is a $N \times N$ matrix and $M$ is a $q \times q$ matrix. The factorization uncouples the physics of the system in two parts, namely intra-sites and inter-sites physics, which we consider in turn:

- **intra-site physics:**

The physics of a given site is fully determined by the states of all cells of that site. If the ratio between the site scale and the cell scale is $10^{-22} / 10^{-35} = 10^{13}$ as it has been suggested above, the number of cells per site is very high. A statistical mechanics treatment is then necessary wherein sites are considered as « macroscopic » systems whose properties are described by order parameters. Let us consider a particular site and let

$$\sigma_\lambda \in \{-1,+1\}$$

be the states of the cells of that site. The order parameter $\varphi$ is defined as the average value of an observable $\chi$:

$$\chi(I) = \frac{1}{q} \sum_\lambda \sigma_\lambda(I)$$

That is:

$$\varphi = \langle \chi \rangle = \frac{1}{Z} \sum_I \chi(I) \rho(I)$$

where the sum is over all possible states $I$ of the system, $\rho(I)$ is the probability for the state $I$ to occur and $Z = \sum_I \rho(I)$ is the partition function. This probability is determined by the functional $H$ (which, hereafter, we call an «energy» in analogy with the usual statistical physics):

$$\rho(I) = \rho(H(I))$$

with:

$$H(I) = -\frac{M}{q} \sum_{\lambda\mu} \sigma_\lambda(I) \sigma_\mu(I)$$

The sum is over *all couples* of cells. The factor $q$ is introduced in the definition of the «energy» so as to make it an extensive quantity and therefore to ensure that the distribution is Maxwellian:

$$\rho(I) = \exp(-\beta H(I))$$

The computation of the order parameter is a classical problem of statistical mechanics. Let us summarize the main steps of this calculation:

One rewrites the average as:
$$\varphi = \frac{1}{Z}\sum_{\chi}\left[\sum_{I(\chi\,\text{given})}\chi(I)\rho(I)\right]$$

Since:
$$\left(\frac{1}{q}\sum_{\lambda}\sigma_{\lambda}\right)\left(\frac{1}{q}\sum_{\mu}\sigma_{\mu}\right) = \frac{1}{q^2}\left(\sum_{\lambda\mu}\sigma_{\lambda}\sigma_{\mu} - q\right) = \chi^2$$

the « energy » may be expressed as:
$$H(I) = -\frac{M}{q}(q^2\chi^2 + q) = H(\chi)$$

and therefore:
$$\varphi = \sum_{\chi}\left(\chi\rho(H(\chi))\left(\sum_{I(\chi\,\text{given})}1\right)\right)$$

One has:
$$\sum_{I,(\chi\,\text{given})}1 = \frac{q!}{\left(\frac{q}{2}(1+\chi)\right)!\left(\frac{q}{2}(1-\chi)\right)!} \approx \exp\left[q\left(\text{Ln}(2) - \left(\frac{1+\chi}{2}\right)\text{Ln}(1+\chi) - \left(\frac{1-\chi}{2}\right)\text{Ln}(1-\chi)\right)\right]$$

by using the Stirling formula whence:
$$\varphi = \frac{1}{Z}\sum_{\chi}\chi\left(\exp\left[q\left(\text{Ln}(2) - \left(\frac{1+\chi}{2}\right)\text{Ln}(1+\chi) - \left(\frac{1-\chi}{2}\right)\text{Ln}(1-\chi)\right) + M\beta(q\chi^2+1)\right]\right)$$

The order parameters are most easily computed by using the following generating function:
$$Z(\beta,h) = \sum_{\{\sigma\}}\exp(-\beta(H(\{\sigma\}) + h\chi))$$

or:
$$Z(\beta,h) = \sum_{\chi}\exp\left[q\left(\text{Ln}(2) - \left(\frac{1+\chi}{2}\right)\text{Ln}(1+\chi) - \left(\frac{1-\chi}{2}\right)\text{Ln}(1-\chi)\right) + M\beta(q\chi^2+1) + h\beta\chi\right]$$

where a linear term $h\chi$ is added to the « energy ». One has:
$$\varphi = \langle\chi\rangle = \lim_{h\to 0}\left(\frac{\partial\text{Ln}(Z(\beta,h))}{\beta\partial h}\right)$$

and:
$$\Delta\varphi^2 = \langle(\chi - \langle\chi\rangle)^2\rangle = \lim_{h\to 0}\left(\frac{\partial^2\text{Ln}(Z(\beta,h))}{\beta^2\partial h^2}\right)$$

One writes the generating function as:
$$Z(\beta,h) = \sum_{\chi}\exp[qF(\chi)]$$

with:
$$F(\chi) = \left(\text{Ln}(2) - \left(\frac{1+\chi}{2}\right)\text{Ln}(1+\chi) - \left(\frac{1-\chi}{2}\right)\text{Ln}(1-\chi)\right) + M\beta\chi^2 + \frac{1}{q}(\beta h\chi + 1)$$

and one defines $\chi_0$ as the value of $\chi$ at the maximum of the function $\left(\dfrac{dF}{d\chi}\right)_{\chi_0} = 0$. Expanding $F$ to second order around the maximum yields:

$$F(\chi) \cong F(\chi_0) + \frac{1}{2}\left(\frac{d^2 F}{d\chi^2}\right)_{\chi_0}(\chi - \chi_0)^2$$

and

$$Z(\beta,h) \cong \sum_\chi \exp(qF(\chi_0)) \cdot \exp\left(q\frac{1}{2}\left(\frac{d^2 F}{d\chi^2}\right)_{\chi_0}(\chi - \chi_0)^2\right)$$

We observes that as soon as $|\chi - \chi_0| > \dfrac{1}{\sqrt{q}}$ the second factor makes the contributions of the various terms to the generating function shrink to zero (one verifies that $\left(\dfrac{d^2 F}{d\chi^2}\right)_{\chi_0} < 0$ indeed). Therefore

$$Z(\beta,h) \cong \exp(qF(\chi_0))$$

The order parameter is $\varphi_0 = \langle \chi \rangle$ which satisfies the equation:

$$\left(\frac{dF}{d\chi}\right)_{\varphi_0} = 0$$

or

$$\frac{1}{2}\operatorname{Ln}\left(\frac{1+\varphi_0}{1-\varphi_0}\right) = \beta M \varphi_0 \qquad (1)$$

Moreover

$$\Delta\varphi^2 = \left\langle (\chi - \langle\chi\rangle)^2 \right\rangle = \langle\chi^2\rangle - \langle\chi\rangle^2 = \lim_{h\to 0}\left(\frac{\partial^2 \operatorname{Ln}(Z(\beta,h))}{\beta^2 \partial h^2}\right) = 0$$

This point is an important one. It shows that the fluctuations of the order parameter vanish whatever the conditions at least as long as the number $q$ of cells per site is large, otherwise the fluctuations are of the order of $1/\sqrt{q}$. In other words, for this model, the mean field theory is an exact theory.

The next point appeals to the Ginzburg criterion: According to this criterion the fluctuations of order parameters of systems with finite connectivities $z$ and dimensionalities $d$ higher or equal to $d = 4$ vanish. For these systems, the domain of parameter $\beta$ where the mean field theory does not apply shrinks to zero accordingly. The properties of fully connected systems are thus similar to those of systems with $d = 4$ and the order parameter $\phi_i$ of a given site $i$ may be considered as a vector determined by its four components:

$$\phi_i = \{\varphi_{i,\alpha}\} \; ; \; i = 1,2,\cdots N \; ; \; \alpha = 0,1,2,3 .$$

The state of a site can be written as:

$$\phi_i = \begin{pmatrix} \varphi_{i0} \\ \varphi_{i1} \\ \varphi_{i2} \\ \varphi_{i3} \end{pmatrix}$$

with $|\phi_i| = \chi_i$ the value of the order parameter at site $i$, and the state of the whole system as a vector $\Phi$:

$$\Phi = \begin{pmatrix} \phi_1 \\ \vdots \\ \phi_i \\ \vdots \end{pmatrix}$$

**- inter-site physics:**

Through the $K$ matrix, the order parameters that develop in the different sites are one another coupled. In terms of states $\Phi$ the «energy» is given by:

$$H = \Phi^T J \Phi$$

with $J = K \otimes G$ and $G$ is $4 \times 4$ matrix.

On the other hand the activity of the system is defined as $A = \Phi^T \Phi$. A dynamics is a state which minimizes the « energy » of the system while keeping its activity. This problem is solved by appealing to the Lagrange parameters technique. One, therefore looks at the minimum of:

$$H - \kappa(\Phi^T \Phi - A) = \Phi^T J \Phi - \kappa(\Phi^T \Phi - A)$$

which must satisfy:

$$\delta \Phi^T (J\Phi - \kappa\Phi) + (\Phi^T J - \kappa\Phi^T)\delta\Phi = 0$$

so that a solution $\Phi$ is an eigenvector of matrix $J$:

$$J\Phi = \kappa\Phi$$

Only positive eigenvalues $\kappa$ can be given a physical meaning. One, therefore may write $\kappa = m^2$ where $m$ is the mass of a particle whose dynamics is given by the corresponding eigenvector:

$$J\Phi = m^2 \Phi$$

The state of minimum energy $\Phi_0$ is a state

$$\Phi_0 = \begin{pmatrix} \phi_1^0 \\ \vdots \\ \phi_i^0 \\ \vdots \\ \phi_N^0 \end{pmatrix}$$

where

$$\phi_i^0 = \begin{pmatrix} \varphi_{i0}^0 \\ \varphi_{i1}^0 \\ \varphi_{i2}^0 \\ \varphi_{i3}^0 \end{pmatrix}$$

and $|\phi_i^0| = \varphi_0$, the solution of equation (1). It is called the (Higgs) vacuum state. Would the parameter $\beta$ be small enough, the vacuum state would be such as $\Phi_0 = 0$ (symmetrical). It is generally admitted that $\beta$ is large enough for the vacuum state to be asymmetrical.

### IV)-Unfolding the physical spaces

So far a physical space is simply a set of points (the sites). The problem of understanding how the usual geometrical properties of physical spaces develop from this set may be related to the problem of understanding how a given transformation of a Lie group develops from the generators of the group.

Let $F(\theta)$ be a transformation of a one parameter Lie group. Since $F(0) = 1$ (the unit transformation) and since the definition of a Lie group implies the transformations to be continuous, an infinitesimal transformation with parameter $\theta/v$ ($v$ large) may be written as:

$$F(\theta/v) \approx 1 - i\frac{\theta}{v}G_\theta$$

where $G_\theta = i\left(\frac{\partial F}{\partial \theta}\right)_{\theta=0}$ is the generator of the group. In the limit $v \to \infty$ the generator may be considered as a property of the geometrical point $\theta = 0$. The transformation $F(\theta)$ is obtained by iterating the infinitesimal transformation. That is to say:

$$F(\theta) = \lim_{v \to \infty}\left(1 - i\frac{\theta}{v}G_\theta\right)^v = \exp(-iG_\theta\theta)$$

The factor i ($i^2 = -1$) makes sure that the transformation is a unitary transformation indeed.

$$F(\theta)F^{-1}(\theta) = F(\theta)F^T(\theta) = \exp(-iG_\theta\theta)\exp(iG_\theta\theta) = 1$$

provided that the generator is hermitian: $G_\theta^* = G_\theta$.

One considers that the order parameters of the cell model of physical systems play the role of generators upon which the physical space develops in the very same way as the Lie transformations develop from the generators of the group. There are two foundamental types of generators: On the one hand there are the localization operators (which allow the definition of positions) and, on the other hand, there are the derivation operators (which allow the definition of momenta). To make them more precise the (symmetrical) matrix $K$ is decomposed according to the LDU (Lower-Diagonal-Upper Matrices) theorem of Banachiewicz[2]:

$$K = D^TUD$$

where $D$ is an $(N \times N)$ upper triangular matrix whose elements are such that:

$$D_{ij} = 0 \text{ for } i > j \text{ (strictly)}$$

$D^T$ is the transposed lower triangular matrix and $U$ is a diagonal matrix. By letting $U^{1/2}D \to D$ this may be rewritten as:

$$K = D^TD$$

One can also argue that, for homogeneous systems, $U$ is a spherical matrix. Then the « energy » is given by:

$$H = \Phi^T\mathbf{D}^T\mathbf{G}\mathbf{D}\Phi$$

where $\mathbf{D} = 1^{(4)} \otimes D$ and $\mathbf{G} = G \otimes 1^{(N)}$. For the sake of hermiticity it is better to rewrite the «energy» as:

$$H = \Phi^T(i\mathbf{D})^T\mathbf{G}(i\mathbf{D})\Phi = \Phi^T(-i\mathbf{D}^T)\mathbf{G}(i\mathbf{D})\Phi$$

Moreover the factorization of the G matrix, if it is possible to have it carried out, allows the dynamics to be put into a simpler form. Let us assume that $\mathbf{G} = \Gamma^T\Gamma$ (where $\Gamma = \gamma \otimes 1^{(N)}$ and $\gamma$ a $4 \times 4$ matrix) then:

$$H = \Phi^T(i\mathbf{D})^T\Gamma^T\Gamma(i\mathbf{D})\Phi = (i\Gamma\mathbf{D}\Phi)^T(i\Gamma\mathbf{D}\Phi)$$

and a solution $\Phi$ of the equation :

$$J\Phi = (i\Gamma D)^T (i\Gamma D)\Phi = m^2 \Phi$$

is also a solution of:

$$(i\Gamma D)\Phi = m\Phi$$

Due to their triangular structure the operators $D$ and $D^T$ allow differential forms to be defined. For example the right position generator is given by:

$$\vec{\Delta} x_\alpha = \sum_j D_{ij} \varphi^0_{j,\alpha}$$

which, for homogeneous systems, is $i$ independant. Similarly the left position generator is given by:

$$\overleftarrow{\Delta} x_\alpha = \sum_j \left(\varphi^0_{j,\alpha}\right)^* \left(D^T\right)_{ji}$$

where $\phi^0_i = \{\varphi^0_{i\alpha}\}$ is the vacuum state. One has: $\left|\vec{\Delta} x_\alpha\right| = \left|\overleftarrow{\Delta} x_\alpha\right| = \Delta x_\alpha$. These generators vanish when the vacuum state is symmetrical (that is to say when $\varphi^0_{i\alpha} = 0$). The $\alpha$ indices are Lorentz indices. The operator:

$$Q_\alpha(X_i) = \exp(-iX_i \Delta x_\alpha)$$

where $X_i$ is the number of iterations of $\Delta x_\alpha$, is a position operator. More precisely $X_i$ is the distance, starting from a site chosen as the origin site and expressed in terms of the standard $\Delta x_\alpha$, along the direction $\alpha$, of a certain site $i$. When all components of the order parameter are taken into account this operator becomes:

$$Q(X_i) = \prod_{\alpha=0}^{3} Q_\alpha(X_i) = \exp\left(-i\sum_{\alpha=0}^{3} X_{i\alpha} \Delta x_\alpha\right) = \exp\left(-i\sum_{\alpha=0}^{3} x_{i\alpha}\right)$$

Differential forms may also be defined for fields. For example:

$$\vec{\Delta} \varphi_{i,\alpha} = i \sum_j D_{ij} \varphi_{j\alpha} = i(D\phi_\alpha)_i$$

is the right differential form for the field $\Phi$. In this expression $\phi_\alpha$ is a vector whose $N$ components are $\varphi_{i\alpha}$. Similarly:

$$\overleftarrow{\Delta} \varphi_{i,\alpha} = -i \sum_j \left(D^T\right)_{ij} \varphi_{j\alpha} = (iD\phi_\alpha)^T_i$$

is the left differential form for the field. Those definitions allow derivation operators to be introduced by letting:

$$\frac{\vec{\Delta} \varphi_{i\alpha}}{\Delta x_\alpha} = i\frac{(D\phi_\alpha)_i}{\Delta x_\alpha} = i\vec{\partial}_\alpha \varphi_i$$

which is to be understood as the right derivative, along the direction $\alpha$, of a scalar field $\phi$ at site $i$. A similar definition is adopted for the left derivative of the field.
By iterating the derivative operator $X_i$ times one obtains the following expression:

$$P_{i\alpha} \phi_\alpha = \exp\left(-iX_i(iD\phi_\alpha)_i\right) = \exp\left(-iX_i \Delta x_\alpha \frac{(iD\phi_\alpha)_i}{\Delta x_\alpha}\right) = \exp\left(-ix_{i\alpha}(i\partial_\alpha \varphi_i)\right)$$

which is really a translation operator since:

$$P_{i\alpha} \phi(x_0) = \exp(x_{i\alpha} \partial_\alpha)\phi(x_0) = \sum_{n=1}^{\infty} \frac{x_{i\alpha}^n}{n!} \partial^n (\phi(x_0)) = \phi(x_0 + x_{i\alpha})$$

according to the formula of Taylor.

Differential forms may be defined for tensorial fields as well. Let $\vartheta_{i,\alpha\beta}$ be the 16 components of a second rank tensor at site $i$. A differential form could be associated to this tensor through the following definition:
$$\Delta\vartheta_{i,\alpha\beta} = i\sum_j D_{ij}\vartheta_{j,\alpha\beta}$$
and the derivative:
$$\frac{\Delta\vartheta_{i,\alpha\beta}}{\Delta x_\alpha} = i\partial_\alpha A_{i,\beta}$$
so introducing a vector field with components $A_\beta$. This definition, however, is not satisfactory because it spoils the hermitian character of the operator as well as the symmetry of indices $\alpha$ and $\beta$. In order to have them restored it is necessary to make the definition symmetrical which leads to:
$$\frac{\Delta\vartheta_{i,\alpha\beta}}{\Delta x} = \frac{\Delta\vartheta_{i,\alpha\beta}}{\Delta x_\alpha} + \left(\frac{\Delta\vartheta_{i,\alpha\beta}}{\Delta x_\beta}\right)^T = i\partial_\alpha A_{i,\beta} + \left(i\partial_\beta A_{i,\alpha}\right)^T = i\left(\partial_\alpha A_{i,\beta} - \partial_\beta A_{i,\alpha}\right)$$
Differential forms, as one can see here, introduce new types of fields. Another example is brought about by the factorization of the G matrix. The product $\Gamma D$ which appears in the dynamics, is expanded according to:
$$\Gamma D = \sum_\mu \Gamma^\mu D_\mu$$
where the set of $\Gamma^\mu$ matrices is a complete set of matrices making the basis of a Clifford algebra (as one shall see below). Differential forms are associated with the $D_\mu$'s such as:
$$\Delta x_\mu^\alpha = \sum_j D_{\mu,ij}\varphi^0_{j\alpha}$$
and
$$\Delta\varphi_{\mu i}^\alpha = i\sum_j D_{\mu,ij}\varphi_{j\alpha}$$
This introduces a new field with $\phi^\alpha$ whose derivatives are given by:
$$\frac{\Delta\varphi_{\mu i}^\alpha}{\Delta x_\mu^\alpha} = i\partial_\mu\phi_i^\alpha$$
This field is called a spinorial field. The $\alpha$ indices are Lorentz indices and the $\mu$ indices are Dirac indices.

The model we have developed so far may be summarized as follows:
The vacuum physical space (meaning a space devoid of excitations) is the space-time space which develops from a series of generators which themselves build up on the vacuum state of sites (meaning the state of lowest energy). This vacuum is necessarily disymmetrical. Particles manifest by modifying the generators so giving rise to a sort of crumpling of the vacuum physical space.

### V)-The Minkowski space

We have seen that the interactions $J_{ij,\alpha\beta}$ are decoupled according to:
$$J_{ij,\alpha\beta} = K_{ij}G_{\alpha\beta}$$

where the interaction matrix $G$ is a $4 \times 4$ symmetrical matrix whose diagonal elements vanish ($G_{\alpha\alpha} = 0$) if the cells do not self-interact. Alls cells of a given site are equivalent and, therefore, the dynamics must be insensitive to a reorganization of cells belonging to one and the same site. This means that $G$ must commute with all ($4 \times 4$) matrices which represent the elements of the symmetrical group $S_4$ of permutations of 4 objects. Let $\Gamma^4$ be this representation. The group $S_4$ has $4! = 24$ elements that are distributed along 5 classes and, consequently it has 5 irreducible representations[4]:

$$\Gamma_1, \Gamma_1^*, \Gamma_2, \Gamma_3, \Gamma_3^*$$

whose orders are 1,1,2,3 and 3 respectively. The table of characters of these representations is the following:

| classes: $l_c$ | (1):1 | (ab):6 | (ab)(cd):3 | (abc):8 | (abcd):6 |
|---|---|---|---|---|---|
| $\Gamma_1$ | 1 | 1 | 1 | 1 | 1 |
| $\Gamma_1^*$ | 1 | -1 | 1 | 1 | -1 |
| $\Gamma_2$ | 2 | 0 | 2 | -1 | 0 |
| $\Gamma_3$ | 3 | 1 | -1 | 0 | -1 |
| $\Gamma_3^*$ | 3 | -1 | -1 | 0 | 1 |

Table 1: Table of characters of group $S_4$
Each class is described by one of its elements (a particular permutation)
$l_c$ is the number of elements of the class.

Taking its table of characters into account:

| classes: $l_c$ | (1):1 | (ab):6 | (ab)(cd):3 | (abc):8 | (abcd):6 |
|---|---|---|---|---|---|
| $\Gamma^4$ | 4 | 2 | 0 | 1 | 0 |

Table 2: Table of characters of representation $\Gamma^4$

the representation $\Gamma^4$ decomposes into irreducible representations according to:

$$\Gamma^4 = \Gamma_1 \oplus \Gamma_3$$

According to the lemma of Schur, the matrix $G$ may be made diagonal accordingly:

$$G = \begin{pmatrix} G_0 & . & . & . \\ . & G_1 & . & . \\ . & . & G_1 & . \\ . & . & . & G_1 \end{pmatrix}$$

On the other hand the trace of $G$, still vanishes and:

$$\text{Tr}(G) = 0 = G_0 + 3G_1$$

Since $G_0 = -3G_1$, $G_0$ et $G_1$ have opposite signs. One lets $|G_0| = 1/c^2 |G_1|$ ($c = 1/\sqrt{3}$ in a natural system of unit where $G_1 = \lambda^2$ and $\lambda$ is determined by the metric scale). $c$, as we see below is the speed of light: One may find satisfactory that the speed of light; which is a physical invariant, is expressed as a dimensionless, and irrationnal, number. Finally the metric tensor g is defined as $g_{\alpha\beta} = \text{sign}(G_\alpha)\delta_{\alpha\beta}$:

$$g = \begin{pmatrix} 1 & . & . & . \\ . & -1 & . & . \\ . & . & -1 & . \\ . & . & . & -1 \end{pmatrix}$$

which is the Minkowski metric tensor. To summarize the physical space is 4-dimensionnal. The 4 dimensions are distributed among three equivalent (space) dimensions on the one hand ($\alpha = 1,2,3$) and one (time) dimension on the other ($\alpha = 0$). $\alpha$ is a Lorentz index.

### VI)-Fermion dynamics

Fermions are fields with components $\varphi_{i,\alpha}$. We have seen that the dynamics of this type of fields is given by:

$$(\Gamma D)\Phi = m\Phi$$

where G has been factorized along $G = \Gamma^T \Gamma$ and $m$ is the fermion mass. To make this expression more explicit we develop the product $\Gamma D$ according to

$$\Gamma D = \sum_\mu \Gamma^\mu D_\mu$$

where $\Gamma^\mu = 1^{(N)} \otimes \gamma^\mu$ and the set of $\gamma^\mu$ is a set of 4 dimensional matrices. The derivation operators $D_\mu$ introduce spinor fields, which, as we have seen above, are defined by the relation:

$$\sum_j D_{\mu,ij} \varphi_{j,\alpha} = i\partial_\mu \phi_i^\alpha$$

$\mu$ is a Dirac index. Developping the equation leads to:

$$\sum_\mu \sum_{jk,\beta\delta} (\Gamma^\mu)_{ij,\alpha\beta} (D_\mu)_{jk,\beta\delta} (\Phi)_{k\delta} = m(\Phi)_{i,\alpha}$$

With:

$$(\Gamma^\mu)_{ij,\alpha\beta} = (\gamma^\mu)_{\alpha\beta} \delta_{ij}$$
$$(D_\mu)_{jk,\beta\delta} = (D_\mu)_{jk} \delta_{\beta\delta}$$

one obtains:

$$\sum_\mu \sum_{k,\beta} (\gamma^\mu)_{\alpha\beta} (D_\mu)_{ik} (\Phi)_{k,\beta} = m(\Phi)_{i,\alpha}$$

or:

$$\sum_\mu \sum_\beta i(\gamma^\mu)_{\alpha\beta} \partial_\mu \phi_i^\beta = m\varphi_{i,\alpha}$$

For the order parameter to be the fermionic field indeed, the right member of this equation must be interpretred as a spinor field too and therefore:

$$\sum_\mu \sum_\beta i(\gamma^\mu)_{\alpha\beta} \partial_\mu \phi^\beta = m\phi^\alpha$$

In a matrix form:

$$\left(\sum_{\mu} i\gamma^{\mu}\partial_{\mu} - m\right)\phi = 0$$

which is the Dirac equation. The properties of Dirac matrices $\gamma^{\mu}$ are determined by the definition of $\Gamma^{\mu}$ matrices. According to the definition:

$$(\Gamma D)^{T}(\Gamma D) = D^{T}GD$$

or:

$$\left(\sum_{\mu}\Gamma^{\mu}\right)^{T}\left(\sum_{\nu}\Gamma^{\nu}\right) = G$$

one has :

$$\frac{1}{2}\sum_{\mu\nu}\left(\gamma^{\mu}\gamma^{\nu} + \gamma^{\nu}\gamma^{\mu}\right) = G$$

which compels the Dirac matrices to obey the following anticommutation equations:

$$\left(\gamma^{\mu}\gamma^{\nu} + \gamma^{\nu}\gamma^{\mu}\right) = 2G_{\mu}1^{(4)}\delta_{\mu\nu}$$

That is to say they form a Clifford algebra. There exists a set of 4 ($4\times 4$) matrices indeed which obey the algebra. In Dirac representation they are given by:

$$\gamma^{\mu} = \begin{pmatrix} 0^{(2)} & \sigma_{\mu} \\ -\sigma_{\mu} & 0^{(2)} \end{pmatrix} \quad \mu = 1,2,3 \quad \gamma^{0} = \begin{pmatrix} 1^{(2)} & 0^{(2)} \\ 0^{(2)} & -1^{(2)} \end{pmatrix}$$

where the 3 ($2\times 2$) $\sigma_{\mu}$ matrices are the Pauli matrices. The first three Dirac matrices are space-like whereas the last one is time-like.

### VII)-Boson dynamics

Let us consider the dynamics of fields that are represented by second rank tensors that is to say fields whose 16 components are $\vartheta_{i,\alpha\beta}$. Their dynamics derives from the following «energy»:

$$H = \sum_{ij,\alpha\beta\gamma\delta} \vartheta_{i,\alpha\beta} J_{ij,\alpha\beta\gamma\delta} \vartheta_{j,\gamma\delta}$$

In this expression the interaction matrix factorizes according to:

$$J_{ij,\alpha\beta\gamma\delta} = K_{ij}(G\otimes G)_{\alpha\beta,\gamma\delta}$$

so that we can write:

$$H = \Theta^{T}D^{T}G^{T}GD\Theta$$

where the state $\Theta$ of the field for the whole system is:

$$\Theta = \begin{pmatrix} \theta_{1} \\ \vdots \\ \theta_{i} \\ \vdots \end{pmatrix}$$

with:

$$\theta_i = \begin{pmatrix} \vartheta_{i,00} \\ \vdots \\ \vartheta_{i,\alpha\beta} \\ \vdots \end{pmatrix}$$

Making the «energy» a minimum quantity, given an activity $A = \Theta^T\Theta$ of the field, leads to the eigenvalue equation:

$$D^T G^T G D \Theta = \kappa \Theta$$

We have seen that the derivative of a second rank tensor may be defined as:

$$((GD)\Theta)_{i,\alpha\beta} = i\sum_\delta \left(G_{\alpha\delta}\partial_\delta A_{i,\beta} - G_{\delta\beta}\partial_\delta A_{i,\alpha}\right) = i\left(G_\alpha \partial_\alpha A_{i,\beta} - G_\beta \partial_\beta A_{i,\alpha}\right)$$

The field $A_{i,\alpha}$, a vector field, is called a gauge field. By letting $\partial^\mu = \sum_\nu G_{\mu\nu}\partial_\nu$ and $\varphi_i^\mu = \sum_\nu G_{\mu\nu}\varphi_{i,\nu}$ the dynamics of the field writes:

$$\sum_\mu \partial^\mu \left(\partial_\mu A_\nu - \partial_\nu A_\mu\right) = m^2 \sum_\mu \vartheta_{\mu\nu}$$

where $m$ is the boson mass. Actually fields described by second rank tensors are boson fields. For massless bosons $m = 0$ the equations reduce to:

$$\sum_\mu \partial^\mu \left(\partial_\mu A_\nu - \partial_\nu A_\mu\right) = 0$$

which are the Maxwell equations.

### VIII)-Gauge interactions

So far we have introduced two types of fields, fermion fields and boson fields. Cross-terms make them to interact. The total «energy» is then given by:

$$H = \Phi^T \Gamma D \Phi + \Theta^T D^T \Gamma^T \Gamma D \Theta + \Phi^T \Omega^T \Theta \Omega \Phi$$

where the first term is the fermionic «energy», the second term the bosonic «energy» and the last term, a term determined by a $(4 \times 4)$ matrix $\Omega$, is the interaction «energy». This term

$$H_{\text{int}} = \sum_{i,\alpha\beta\gamma\delta} \varphi^*_{i,\alpha} \Omega^T_{i,\alpha\beta} \Theta_{i,\beta\gamma} \Omega_{i,\gamma\delta} \varphi_{i,\delta}$$

is rewritten as:

$$H_{\text{int}} = \sum_{i,\alpha\beta\gamma\delta} \varphi^*_{i,\alpha} \varphi_{i,\beta} \Xi_{i,\alpha\beta\gamma\delta} \Theta_{i,\gamma\delta} = \Phi^T \Phi \Xi \Theta$$

It is a purely local expression (since the D matrices do not come into play) and the characteristics of the fermion-boson interaction are fully determined by the $16 \times 16$ matrix $\Xi$. To make these properties more precise we assume that the interaction «energy» is gauge invariant which means that it remains unchanged under internal site symmetries that is to say under permutation symmetries $P$. One therefore looks for $(16 \times 16)$ matrices $W(P)$ which make a representation of group $P_4$ while leaving the matrix $\Xi$ unchanged. A convenient representation is obtained by considering the direct product of matrices associated with the symmetries of the boson field $\Theta$ on the one hand with those associated with the fermion field $\Phi$ on the other. The first set, which describes the transformations of the boson field, is the set of the 24 $(4 \times 4)$ matrices $U(P)$ representating the permutations of 4 objects.

$$U(P = 1234 \to 3142) = \begin{pmatrix} . & . & 1 & . \\ 1 & . & . & . \\ . & . & . & 1 \\ . & 1 & . & . \end{pmatrix}$$

is an example. The boson fields are Lorentz invariant indeed as are, for example, the Maxwell equations. In section V we called this representation $\Gamma^4$. It cannot be used to describe the transformations of the fermion field, however. The matrix G transforms according to this representation but not the matrices $\Gamma$ which result from its factorization (it is reminded that $G = \Gamma^T\Gamma$). This fact is in a conspicuous position when considering the Dirac matrices $\gamma^\mu$. The application of one of the matrix of $\Gamma^4$ to a bi-spinor field mixes the components of the bi-spinor which then looses its character. To transform bi-spinors into bi-spinors the representation matrices must be of the form:

$$V(P) = \begin{pmatrix} \omega(P) & . \\ . & \varpi(P) \end{pmatrix}$$

where $\omega(P)$ is a $(2 \times 2)$ representation of group $P_4$. We have seen that a second order irreducible representation of $P_4$, named $\Gamma^2$, does exist and therefore it is possible to choose the representation for the transformation of the fermion field as:

$$V(P) = \Gamma^2(P) \otimes 1^{(2)} = \begin{pmatrix} \Gamma^2(P) & . \\ . & \Gamma^2(P) \end{pmatrix}$$

The representation associated with $\Xi$ is therefore such as:
$$W(P) = U(P) \otimes V(P)$$
and one looks for its decomposition into the irreducible representations of $P_4$. The $(2 \times 2)$ matrices $\Gamma^2(P)$ may be expressed in terms of the unit matrix $\sigma_0 = 1^{(2)}$ and the 3 Pauli matrices $\sigma_\alpha$; $\alpha = 1,2,3$:

$$\Gamma^2(P) = \sum_{\alpha=0}^{3} g_\alpha(P)\sigma_\alpha$$

If the interaction driven by $\Gamma^2(P)$ is to obey Lorentz invariance, the parameters $g_\alpha$ are such that:

$$g_0 = Ag' \; ; \; g_1 = g_2 = g_3 = Ag \text{ with}$$
$$\frac{g'}{g} = \frac{g_0}{g_1} \approx \sqrt{\left|\frac{G_1}{G_0}\right|} = c$$

where $A$ is some constant and where the notations $g$ and $g'$ have been introduced for the sake of using the notations of electro-weak theories [6]. Then one may write:
$$\Gamma^2 = g_0\Gamma_1 + g_1\Gamma_2{}'$$
($\Gamma_1$ is a $2 \times 2$ unit matrix and $\Gamma_2{}'$ is a $2 \times 2$ traceless matrix) or:
$$\Gamma^2(P) = g'A_0\sigma_0 + g\vec{A}.\vec{\sigma}$$
and

$$W(P) = \Gamma^4(P) \otimes \begin{pmatrix} g_0\Gamma_1 + g_1\Gamma_2{}' & . \\ . & g_0\Gamma_1 + g_1\Gamma_2{}' \end{pmatrix} = \begin{pmatrix} \Gamma^1 & . \\ . & \Gamma^3 \end{pmatrix} \otimes \begin{pmatrix} g_0\Gamma_1 + g_1\Gamma_2{}' & . \\ . & g_0\Gamma_1 + g_1\Gamma_2{}' \end{pmatrix}$$

or:

$$W = \begin{pmatrix} \Gamma^1 \otimes (g_0\Gamma_1 + g_1\Gamma_2') & . & . & . \\ . & \Gamma^3(P) \otimes (g_0\Gamma_1 + g_1\Gamma_2') & . & . \\ . & . & \Gamma^1 \otimes (g_0\Gamma_1 + g_1\Gamma_2') & . \\ . & . & . & \Gamma^3(P) \otimes (g_0\Gamma_1 + g_1\Gamma_2') \end{pmatrix}$$

This $(16 \times 16)$ representation $W$ therefore decomposes into 4 blocks of sizes 2,6,2 and 6 respectively. It can be considered as a direct sum of two $(8 \times 8)$ matrices such as:

$$\begin{pmatrix} \Gamma^1 \otimes (g_0\Gamma_1 + g_1\Gamma_2') & . \\ . & \Gamma^3 \otimes (g_0\Gamma_1 + g_1\Gamma_2') \end{pmatrix}$$

It is to be reminded that the $\Gamma^4$ representation determines the metrics of the space according to:

$$\Gamma^4 \rightarrow \begin{pmatrix} G_0 & . & . & . \\ . & G_1 & . & . \\ . & . & G_1 & . \\ . & . & . & G_1 \end{pmatrix}$$

with $G_1 = -\frac{1}{3}G_0$. Therefore the interaction matrix takes the following form:

$$\begin{pmatrix} G_0(g_0\Gamma_1 + g_1\Gamma_2') & . \\ . & -\frac{G_0}{3} 1^{(3)} \otimes (g_0\Gamma_1 + g_1\Gamma_2') \end{pmatrix}$$

This set of interactions gives labels to particles.

The first block could be associated with particles called the leptons. The basis states which span this 2 dimensional subspace are labelled by a bivalued quantum number, the isospin quantum number. They define two sorts of leptons. One of the isospin quantum numbers, $\sigma_3 = -1/2$, labels the electron-type leptons and the other, $\sigma_3 = +1/2$ labels the neutrino-type leptons. Similarly a bivalued isospin quantum number is attached to the second block each determining a 3 fold sub space. This block could be associated with quark particles. The first isospin quantum number $\sigma_3 = -1/2$ labels the (down-type) d-quark and the other $\sigma_3 = +1/2$ the (up-type) u-quark. The three fold degeneracy of each of these subspaces is, of course, interpreted as colour quantum numbers.

Let us now consider the electromagnetic properties of these particles. We have seen that the first block may be written as:

$$W = g'A_0\sigma_0 + g\vec{A}.\vec{\sigma}$$

More precisely when developped, the symmetry operation takes the form:

$$W = \exp(ig'YA_0 + ig\vec{A}.\vec{\sigma}) \qquad (2)$$

an element of Lie group $U(1) \otimes SU(2)$. We observe that this factorization is a direct (and necessary) consequence of the partitionning of parameters $g_\alpha$ into three identical parameters $g$ on the one hand and one parameter $g'$ on the other. $Y$ is the hypercharge. This makes the connection with the usual derivation of the electroweak interaction formalism .

The present approach, however, makes it possible to go a bit farther since we have a relation between the parameters $g$ and $g'$:

$$\frac{g'}{g} = c$$

According to the usual reasonning, which we reproduce below, we consider the covariant derivative which writes:

$$D_\mu = \partial_\mu + ig'YA_{\mu 0} + ig\vec{A}_\mu.\vec{\sigma} = \partial_\mu + ig\left(cYA_{\mu,0}1^{(2)} + \left(A_{\mu,1}\sigma_1 + A_{\mu,2}\sigma_2 + A_{\mu,3}\sigma_3\right)\right)$$

Introduced in the Lagrangian

$$L = \sum_\mu \left(D^\mu \phi_0\right)*\left(D_\mu \phi_0\right)$$

where $\phi_0$ is the disymmetrical vacuum state the covariant derivative yields:

$$L = \sum_\mu \phi_0 * \left(\partial_\mu - ig(cA_{\mu,0}\sigma_0 + \vec{A}_\mu.\vec{\sigma})\right)\left(\partial_\mu + ig(cA_{\mu,0}\sigma_0 + \vec{A}_\mu.\vec{\sigma})\right)\phi_0$$

with $\phi_0 = \begin{pmatrix} 0 \\ v \end{pmatrix}$. The $v^2$ term, the relevant one, is:

$$v^2 g^2 \left| \begin{matrix} A_{\mu,1} - iA_{\mu,2} \\ -A_{\mu,3} + cA_{\mu,0} \end{matrix} \right|^2$$

since:

$$\vec{A}_\mu.\vec{\sigma} = \begin{pmatrix} A_{\mu,3} & A_{\mu,1} - iA_{\mu,2} \\ A_{\mu,1} + iA_{\mu,2} & -A_{\mu,3} \end{pmatrix}$$

By defining:

$$W_\mu^+ = \frac{1}{\sqrt{2}}\left(A_{\mu,1} + iA_{\mu,2}\right)$$

$$W_\mu^- = \frac{1}{\sqrt{2}}\left(A_{\mu,1} - iA_{\mu,2}\right)$$

this term becomes:

$$v^2 g^2 \left[ \left(|W_\mu^+|^2 + |W_\mu^-|^2\right) + \left(A_{\mu,3} \quad A_{\mu,0}\right)\begin{pmatrix} 1 & -c \\ -c & 1 \end{pmatrix}\begin{pmatrix} A_{\mu,3} \\ A_{\mu,0} \end{pmatrix} \right]$$

The last contribution is made diagonal by letting:

$$B_\mu = A_{\mu,0}\cos\theta_W - A_{\mu,3}\sin\theta_W$$

$$Z_{0,\mu} = A_{\mu,0}\sin\theta_W + A_{\mu,3}\cos\theta_W$$

which are the eigenvectors of the matrix provided that:

$$\sin\theta_W = \frac{c}{\sqrt{1+c^2}}$$

$$\cos\theta_W = \frac{1}{\sqrt{1+c^2}}$$

$$\text{tg}\theta_W = c$$

($W_\mu^\pm$ forward the interaction through charged currents. $Z_{0\mu}$ forwards the interaction through neutal currents and $B_\mu$ forwards the electromagnetic interactions (the photon)). One of the eigenvalue vanishes as it must be for the photons to be massless. The angle $\theta_W$ is the Weinberg angle. In natural units we have seen that, $c = 1/\sqrt{3}$ and, therefore:

$$\theta_W = \frac{\pi}{6}.$$

Finally:

$$\sin^2\theta_W = \frac{c^2}{1+c^2} = \frac{1}{4} = 0.25$$

a value which is close, but not really equal, to the experimental value of $\sin^2\theta_W = 0.23$.

The hypercharge $Y$ (equation 2) is connected to the electric charge $Q$ by the following relation:

$$Q = \frac{Y}{2} + \sigma_3$$

The electric charge of electronic leptons (where $Y = -1$) is therefore:

$$Q_e = -\frac{1}{2} - \frac{1}{2} = -1$$

whereas the electric charge of neutrino leptons is:

$$Q_\nu = -\frac{1}{2} + \frac{1}{2} = 0.$$

The derivation of electric charges of quarks is identical except that the parameter $G_0$ becomes $G_0 \to -\frac{1}{3}G_0$ and therefore $Y \to -\frac{1}{3}Y = \frac{1}{3}$. One concludes that the electric charge of down quarks is:

$$Q_d = \frac{1}{6} - \frac{1}{2} = -\frac{1}{3}$$

and the electric charge of up quarks is:

$$Q_u = \frac{1}{6} + \frac{1}{2} = \frac{2}{3}$$

The model, therefore, seems to form a convenient framework for the description of the Standard Model. It must be stressed, however, that it shows some short-comings. For example it misses an essential ingredient: the parity non-conservation. On the other hand the particle masses cannot be computed without any further information regarding the interaction matrix K.

### IX)-Canonical quantization and path integrals

The fields whose dynamics have been derived in the preceeding sections are quantized fields. This statement has to be proven. Canonical quantization is the quantization procedure which fits best the present approach. As we know the canonical quantization is associated to the properties of position and momentum operators. One therefore strives to compute the commutator $[X, iD]$ of these two operators. On the basis which makes diagonal the position operator, the matrix element $(XD)_{ij}$ is given by:

$$(XD)_{ij} = X_i \Delta x \frac{D_{ij}}{\Delta x} = X_i D_{ij}$$

Also:

$$(DX)_{ij} = X_j D_{ij}$$

and therefore:

$$([X, iD])_{ij} = i(X_i - X_j)D_{ij}$$

Applying the commutator to a state $\Phi$ whose components are $\phi_k$ yields:

$$([X, iD]\Phi)_k = i\sum_l [X, D]_{kl} \phi_l = i\sum_l (X_k - X_l) D_{kl} \phi_l$$

The integrant factor is non-vanishing for $k \cong l$ and one may write:
$$([X, iD]\Phi)_k \cong i\sum_l (X_k - X_l) D_{kl} \phi_k = i\hbar \phi_k$$

or:
$$[X, iD] = [x, i\partial] = i\hbar$$

which is the canonical quantization rule indeed. This derivation gives a value to the Planck constant $\hbar$:
$$\hbar = \sum_l (X_k - X_l) D_{kl} = \sum_l |X_l D_{0l}|$$

where the site $k = 0$ is taken as the origin site. The properties of vacuum enters this expression since $X_l = x_l / \Delta x$ with $\Delta x = \left|\sum_j D_{ij} \varphi_j^0\right|$.

More generally the model allows any aspect of the usual formalism of quantum mechanics to be recovered. The path integral of Feynman is an example. Let us consider two states $\Phi_a$ et $\Phi_b$. The quantity $\Phi_b^T \Phi_a$ is the correlation (or Green function) between the two states. The correlation may be computed by introducing the unit operator $\sum_k \Phi_k \Phi_k^T = 1$ as many times as necessary leading to:
$$\Phi_b^T \Phi_a = \sum_{klm\cdots} \Phi_b^T \Phi_k \Phi_k^T \Phi_l \Phi_l^T \Phi_m \Phi_m^T \cdots \Phi_a$$

The summation can be carried out either by considering all indices in turn or by selecting a given permutation of indices and summing over all permutations afterwards. A specific permutation is called a path « $C$ ». With this last procedure the correlation function takes the form of a path summation:
$$\Phi_b^T \Phi_a = \sum_C \Phi_b^T \Phi_{k(c)} \Phi_{k(c)}^T \Phi_{l(c)} \Phi_{l(c)}^T \Phi_{m(c)} \Phi_{m(c)}^T \cdots \Phi_a$$

Let us consider the correlation $\Phi_{k(c)}^T \Phi_{l(c)}$. Two states $\Phi_k$ et $\Phi_l$ are close to one another if $\Phi_k = \Phi_l + \partial \Phi_l$ that is to say if $\varphi_{i,\alpha}^k = \varphi_{i,\alpha}^l - i\sum_{j,\delta} D_{ij} \gamma_{\alpha\delta} \varphi_{j,\delta}^l$. Therefore the correlation function between two close states is given by:
$$\Phi_k^T \Phi_l = 1 + \Phi_l^T \partial \Phi_l = 1 - i\sum_{ij,\alpha\beta\delta} D_{ij} \gamma_{\alpha\delta} \varphi_{i,\delta}^{*l} \varphi_{j,\delta}^l \approx 1 - i\sum_{ij,\alpha\beta\delta} D_{ij} \gamma_{\alpha\delta} \varphi_{i,\delta}^{*k} \varphi_{j,\delta}^k = 1 - iL(k)$$

$L(k) = \Phi_k^T \Gamma D \Phi_k$ is called the Lagrangian of the system in state $k$. This is the generator of the correlation function which therefore writes:
$$\Phi_k^T \Phi_l = \exp(-iL(k))$$

As a whole
$$\Phi_b^T \Phi_a = \sum_C \prod_k \exp(-iL(k(C)) = \sum_{C(a\to b)} \exp\left(-i\sum_k L(k(C))\right) = \sum_{C(a\to b)} \exp(-iS(C))$$

$S = \sum_k L(k(C))$ is called the Action, a quantity associated with a given path $C$ and the correlation function takes the form of a path integral.

### X)- Bounded metrics spaces and general relativity

The energy of a physical space devoid of any excitation is given by:
$$H_0 = \Phi_0^T J \Phi_0 = \Phi_0 * D^T G D \Phi_0$$

Explicitely:
$$H_0 = \sum_{ijkl,\alpha\beta\gamma\delta} \varphi_{i\alpha}^0 (D^T)_{ij,\alpha\beta} (G)_{jk,\beta\gamma} (D)_{kl,\gamma\delta} \varphi_{l\beta}^0$$

With:
$$(D^T)_{ij,\alpha\beta} = (D^T)_{ij} \delta_{\alpha\beta}$$
$$(G)_{jk,\beta\gamma} = (G)_{\beta\gamma} \delta_{jk}$$
$$(D)_{kl,\gamma\delta} = (D)_{kl} \delta_{\gamma\delta}$$

the energy writes:
$$H = \sum_{ijk,\alpha\beta} \varphi_{i\alpha}^0 (D^T)_{ij} (G)_{\alpha\beta} (D)_{jk} \varphi_{j\beta}^0$$

By introducing the differential forms we have defined above:
$$dx_\alpha(j) = \sum_i \varphi_{i\alpha}^0 (D^T)_{ij}$$
$$dx_\beta(j) = \sum_k (D)_{jk} \varphi_{j\beta}^0$$

we obtain:
$$H_0 = \sum_j \left( \sum_{\alpha\beta} dx_\alpha(j) G_{\alpha\beta} dx_\beta(j) \right)$$

Actually the order parameter present (weak) fluctuations. The *G* matrix is site dependent and the energy is then to be written as:
$$H_0 = \sum_j \left( \sum_{\alpha\beta} dx_\alpha(j) G_{\alpha\beta}(j) dx_\beta(j) \right)$$

which, in the limit of continuous spaces, identifies with the usual general relativity expression:
$$H = \int \sum_{\mu\nu} g_{\mu\nu}(x) dx_\mu dx_\nu$$

In the model we put forward in this article the origin of space-time curvatures is to be found in the fluctuations of the order parameters which determine the metrics of space-time. A similar view is put forward by Moffat[7]. When fluctuations are ignored the metric tensor is constant and no curvature shows up. Assuming that the fluctuations are very large would lead to random metrics with the loss of all geometrical properties. We are therefore led to the conclusion that the fluctuations, although non zero, are extremely weak which amounts to saying that the number of cells per site is very large (since the fluctuations are proportionnal to the inverse square root $1/\sqrt{q}$ of the number of cells ). This also implies that the parameter $\beta$, which plays the role of the inverse of a temperature, is large enough, a condition which is also necessary for an asymmetric vacuum state to build up.

The fluctuations of the metric tensor modify the dynamics of particles in a way which is easy to understand. For the fermion dynamics we considered the following factorization:
$$JG = (\Gamma D)^T (\Gamma D)$$

then we expanded the product $(\Gamma D)$ on a basis whose size (*m*) is, for the time being, left undetermined.

$$\Gamma D = i \sum_{\mu=1}^{n} \Gamma^{\mu} D_{\mu}$$

$n$ is the number of conditions the correlations have to satisfy and:

$$\Gamma^{\mu} = 1^{(N)} \otimes \gamma^{\mu}$$

We have seen that the $m \times m$ matrices $\gamma^{\mu}$ must obey the following anticommutation conditions:

$$(\gamma^{\mu}\gamma^{\nu} + \gamma^{\nu}\gamma^{\mu}) = 2G_{\mu\nu} 1^{(m)}$$

When fluctuations are ignored the metric tensor is diagonal and the number of conditions to be satisfied is $n = 4$ which is the number of $\gamma^{\mu}$ matrices. The problem is that of choosing the size $m$ of matrices. $m=2$ is not satisfactory because it is not possible to find 4 order 2 matrices which obey the anticommutation rules. This is possible when $m=4$: There exist 4 order 4 matrices (the Dirac matrices) which fulfill the anticommutation rules (the Clifford algebra). When fluctuations have to be taken into account the number of rules to be satisfied is $n=10$ (assuming that the fluctuation tensor is symmetrical $G_{\mu\nu} = G_{\nu\mu}$). The independent components are the 4 diagonal elements and 6 non-diagonal elements so giving 6+4=10 independent elements. One then needs 10 generators (instead of 4) for the dynamics of fermions to be determined. The same sort of reasonning may be applied to boson dynamics. The difference is that one has now to consider the fluctuations of the $(16 \times 16)$ matrix $\Xi$. This gives rise, in principle, to 256 correlation functions and as many conditions to satisfy. In reality this number is much less: There are 16 diagonal components and 240 off-diagonal components. Since indices permutations yield the same correlation functions and since the number of permutations of indices is 4!=24, the number of independent components reduces to 16+240/24=26. It is worth noting that 10 and 26 are precisely the number of dimensions that superstring theories assign to fermion and boson descriptions respectively.

The Dirac matrices, as we know, are obtained from the 3 Pauli matrices. Explicitely:

$$\gamma_i = \begin{pmatrix} 0^{(2)} & \sigma_i \\ -\sigma_i & 0^{(2)} \end{pmatrix} ; i = 1,2,3 \quad \gamma_0 = \begin{pmatrix} 1^{(2)} & 0^{(2)} \\ 0^{(2)} & -1^{(2)} \end{pmatrix}$$

which verify the anticommutation rules

$$\gamma_{\mu}\gamma_{\nu} + \gamma_{\nu}\gamma_{\mu} = 2G_{\mu\nu}\delta_{\mu\nu} \quad \mu,\nu = 0,1,2,3$$

as they have to when the fluctuations vanish.

Introducing the fluctuations makes it necessary for $n=10$ order $m$ matrices $\xi_{\mu}$ to be defined. A possible choice with $m=4$ is as follows:

-the 4 Dirac matrices:

$$\xi_0 = \begin{pmatrix} 1 & 0 \\ 0 & -1 \end{pmatrix} ; \quad \xi_i = \begin{pmatrix} 0 & \sigma_i \\ -\sigma_i & 0 \end{pmatrix} \quad i = k = 1,2,3$$

-3 diagonal matrices:

$$\xi_k = \begin{pmatrix} \sigma_i & 0 \\ 0 & -\sigma_i \end{pmatrix} \quad i = 1,2,3 \; ; \; k = 4,5,6$$

-and finally 3 symmetrical matrices:

$$\xi_k = \begin{pmatrix} 0 & \sigma_i \\ \sigma_i & 0 \end{pmatrix} \quad i = 1,2,3 \; ; \; k = 7,8,9$$

These are independent, although non orthogonal, matrices. In the eigenvalue equation:

$$D^T G D \Phi = (D^T \Gamma^T)(\Gamma D)\Phi = \kappa \Phi$$

the $\Gamma$ matrix is expanded on the $\xi_k$ basis:
$$\Gamma = i \sum_{k=0,1,\ldots,9} \eta_k \xi_k$$

Identifying:
$$\left(\sum_k \eta_k \xi_k\right)\left(\sum_l \eta_l \xi_l\right) = \sum_{kl} \eta_k \eta_l (\xi_k \xi_l + \xi_l \xi_k) \equiv G$$

gives, in principle, solutions for the 10 unknown quantities $\eta_k$ given the 10 parameters $G_{\alpha\beta}$. More precisely the anticommutators are expanded on the basis:
$$(\xi_k \xi_l + \xi_l \xi_k) = \sum_m C_{kl}^m \xi_m$$

where the coefficients $C_{kl}^m$ are structure parameters and the equations to be solved are the following (equations (2)):
$$\sum_{kl}\left(\sum_m C_{kl}^m (\xi_m)_{\alpha\beta}\right) \eta_k \eta_l = G_{\alpha\beta} \qquad (2).$$

whose solutions, provided they do exist, are $\eta_\mu = \eta_\mu(G)$. The dynamics is then given by:
$$(\Gamma D)\Phi = \sqrt{\kappa}\Phi$$

Explicitely:
$$\sum_{j\beta}(\Gamma D)_{ij,\alpha\beta}(\Phi)_{j\beta} = \sum_{\mu=0}^{9}\left(\sum_{jk,\beta\gamma}(i\eta_\mu \Gamma^\mu)_{ij,\alpha\beta}(D_\mu)_{jk,\beta\gamma}\right)(\Phi)_{k\gamma} = \sqrt{\kappa}(\Phi)_{i\alpha}$$

With:
$$(\Gamma^\mu)_{ij,\alpha\beta} = (\xi^\mu)_{\alpha\beta} \delta_{ij}$$
$$(D_\mu)_{jk,\beta\gamma} = (D_\mu)_{jk} \delta_{\beta\gamma}$$

one obtains:
$$\sum_\mu\left(\sum_{j,\beta}(i\eta_\mu \xi^\mu)_{\alpha\beta}(D_\mu)_{ij}\right)(\Phi)_{j,\beta} = \sqrt{\kappa}(\Phi)_{i\alpha}$$

equations wherein our definition of derivatives
$$\sum_j (D_\mu)_{ij}(\Phi)_{j\beta} = \partial_\mu(\phi_i^\beta)$$

are introduced. $\phi_i^\beta$ is the component of the spinor $\beta$ which describes the fermion field at site $i$. Since $(\Phi)_{i,\alpha} \equiv \phi_i^\alpha$ and $\sqrt{\kappa} = m$ the dynamics is finally given by:
$$\left(i\hbar \sum_\mu \eta_\mu(G) \xi^\mu \partial_\mu - mc^2\right)\phi(x) = 0$$

which is known as quantum dynamics in curved spaces. The associated Lagrangian is
$$L(x) = \phi^*(x)\left(i\hbar \sum_\mu \eta_\mu(G) \xi^\mu \partial_\mu - mc^2\right)\phi(x)$$

When fluctuations vanish $\eta_\mu = 1$; for $\mu = 0,1,2,3$ $\eta_\mu = 0$; for $\mu = 4,\ldots,9$. When they do not vanish one can rewrite the Lagrangian as:

$$L(x) = \phi^*(x)\left[\left(i\hbar\sum_{\mu=0}^{3}\gamma^\mu\partial_\mu - mc^2\right) + i\hbar\sum_{\mu=0}^{9}\eta_\mu(G)\xi^\mu\partial_\mu\right]\phi(x)$$

$$= \phi^*(x)\left(i\hbar\sum_{\mu=0}^{3}\gamma^\mu\partial_\mu - mc^2\right)\phi(x) + \phi^*(x)\left(i\hbar\sum_{\mu=0}^{9}\eta_\mu(G)\xi^\mu\partial_\mu\right)\phi(x) = L_D + L_{int}$$

where the quantities $\eta_\mu$ now represent the deviations of the metric tensor with respect to the unperturbed tensor. We observe that the Lagrangian is made of two parts: on the one hand there is the usual Lagrangian which describes the dynamics of free fermions and, on the other, there is an interaction Lagrangian between the fermion field and a 10 components tensor field which we interpret as the gravitationnal field. Since there are no $(\eta_\mu)^2$ quadratic terms this field is massless and its propagator varies as $1/r$. Introducing normalized fluctuations $\langle\eta_\mu\rangle$ one writes:

$$\eta_\mu = \mu\langle\eta_\mu\rangle$$

where the parameter $\mu$ is proportionnal to the mass of the particle. From:

$$(\Gamma D)\Phi = m\Phi$$

one has:

$$m = \frac{\Phi^*(\Gamma D)\Phi}{\Phi^*\Phi}$$

and therefore $\Gamma \propto \mu \propto m$ indeed. One may then let:

$$\mu = m\sqrt{G_g}$$

where $G_g$ is the gravitation constant. To the lowest order of perturbation theory, the interaction between two fermions with masses $m$ and $m'$ is given by:

$$V(r) = m\sqrt{G_g}\frac{1}{r}m'\sqrt{G_g} = \frac{G_g\,mm'}{r}$$

which is the gravitationnal interaction. Since the fluctuations are assumed very weak, the gravitationnal interaction may be many orders of magnitude smaller than the other interactions.

Solving equations (2) is solving a set of 10 quadratic equations. According to the theorem of Bezout the system generally has $2^{10}$ solutions. All solutions cannot be accepted however, and many are degenerated. For example if $\{\eta_\mu\}$ is a solution, $\{-\eta_\mu\}$ is also a solution. It must be realized that a particular gravitationnal constant is, in principle, to be associated with every solution. This may be interpreted as if all particles experience the same gravitationnal field but have their masses changed by the effect of fluctuations (one keeps the gravitation constant while changing the masses instead of changing the gravitation constant while keeping the masses). This could be an explanation for the existence of the 3 families of particles which have exactly the same properties except as regard their masses.

Here is the place where to make some comments regarding the problem of dimensionality. We have argued that the minimal dimensionality of the physical space is $d = 4$ but one can wonder why not choosing another, higher, dimensionality. For example we could have considered the five-dimensional space of Klein and Kaluza. The basic group is then the symmetrical group $S_5$ of the permutations of five objects and we must reduce a $\Gamma^5$ representation of the group according to its irreducible representations. We find that $\Gamma^5$ decomposes according to two one-dimensional representations on the one hand and one three-dimensional representation on the other. This leads to a space endowed with a three degenerate space dimensions and two time dimensions so introducing two different sorts of light velocities. This space does not fit our conventional space-time physical space. A four-dimensional space is enough to describe the properties of vacuum (the ground state of the physical space), as far as fluctuations of the metric tensor are negligible. When they are not, the description of physical phenomena may demand extra-dimensions which can be considered as spanning the internal space of relevant particles. These spaces are linked to the very existence of the particles and disappear when they are not excited. The point of view which is put forward in this article is that dimensions are not intrinsic properties but that they entirely depend on the nature of particles they house.

## XI)-Summary, discussion, and conclusion

The synthetic view of physics we have nowadays owes much to the existence of characteristic energy scales which determine specific types of particles. Although it is very tempting to carry on in that direction and to imagine that quarks are themselves built on some sort of more foundamental particles it seems that present speculations, for example those which are discussed in string theories, are more focused on the properties of the physical space itself. The present approach borrows to both points of view: Beyond the characteristic energy scale which is associated to quarks a vast domain of energy would exist whose foundamental entities would be « cells » with each cell carrying an elementary bit of information. This domain would extend up to the Planck energy. The cells interact and these are the interactions which build up all the geometrical and dynamical properties of the physical space. All physical information is thus contained in these interactions, a point which we think we have made clear enough when expressing the three foundamental constants the Planck length is made of, namely the speed of light, the Planck constant and the gravitationnal constant in terms of these interactions. Obviously a significant step would be carried out if some general principle would make us enable to determine the interactions. All what we did so far has been replacing a set of constants by another, may be larger, set of parameters.

For the time being let us gather the various features whose understanding may be made easier by the present model.
-The physical space is a 4 dimensional space.
-The dimensions are divided up into three equivalent space dimensions and one time dimension.
-The metrics is the Minkowski metrics.
-The vacuum state is disymmetrical and disymmetry is necessary a condition for the various dimensions to unfold.
-The physical space excitations may be viewed as local deformations of the metrics.
-The dynamics of vector type order parameters is determined by Dirac equations.
-The dynamics of tensor type order parameters is, at least for massless excitations, determined by Maxwell equations.
-The interactions between these two types of excitations are divided up in, and only in, three types of interactions, namely in order 1 (e.m.), 2 (weak), and 3 (strong) interactions.
-More generally the model allows the Standard Model to be derived. The formalism of electroweak interactions is recovered and yields, for the Weinberg angle, a value of:

$\sin^2(\theta_W) = 0.25$. One also has an explanation for the quantization of electric charges as they are observed.
-The origin of gravitationnal interactions is to be found in the fluctuations of the metric tensor.

It must be stressed that the model is not fully satisfactory particularly as regards the parity non-conservation properties. On the other hand one could appeal to the central hypothesis of the model, namely the existence of a metric scale, to consider some problems pending in modern physics with a fresh eye. We end this article with these considerations.

i)-The first problem is related to the existence of all sorts of infinite quantities in theoretical physics, especially in perturbation theories. Those infinite quantities mainly arise from the behaviour of high energy spectra of free particles. The existence of a metric scale would make the spectra sharply diverge at the metric scale so introducing cut-offs in pathological integrals. Of course this is the easiest way to solve the problem of infinities but without any incentive to introducing cut-offs, the theoretical physicists prefer to appeal to much more handsome techniques such as dimensionnal renormalization. One of the main results of renormalization theories is that the interaction parameters are energy dependent. For example the fine structure constant increases from $\alpha = 1/137$ for low energies to $\alpha = 1/128$ in the range of electroweak interaction energies $\approx 80\,GeV$, which is observed. One could say that if renormalization theories are essentially correct in domains far from the metric scale, they must loose their relevance in domains closer to the metric scale. Moreover the metric scale makes it impossible for excitations whose wave lengths are less than this scale, to exist. Any attempt at creating such an excitation would result in its decomposition in a number of less energetic excitations.

ii)-The second issue relates to the problem of measurement and, more specifically, to the EPR paradox whose relevance has been clearly established in the A. Aspect experiments[8].
Let us remember that the A. Aspect experiments refer to the observation of polarization correlation between two simultaneously created photons and that they show that correlation is still observed when the polarization of one photon is modified eventhough the distance between the two analysers is such that they cannot be causaly related. The central idea of a metric scale based explanation is that the two photons are created in *one and the same site*. More precisely the state of cells in this site is a two virtual photons state. The site, however, is not still integrated to the experimental environment: it is everywhere and nowhere. The integration is induced by an analyser. At this very moment the creation site splits in two distinct (detected) sites but since the distances between the cells of the creation site were all zero, information regarding the polarization states are readily transmitted to the two detected sites. It is to be stressed that this mechanism does not appeal to extra variables and, therefore, this model is not a hidden variable model. This way of explaining the Aspect experiments looks a bit strange but it can be admitted that is is by no means stranger than the many universe hypothesis of Everett[9].

iii)-The last problem we address is that of the homogeneity of Universe. This homogeneity seems to contradict the causality principles of relativity at least in early Universe since homogeneity makes it necessary for matter end energy to instantaneously spread into the whole system. This has been explained by appealing to the so-called inflationary model of Guth. The central role of this model is played by the cosmological constant $\Lambda$ of Einstein:

This term would have induced a strong repulsion during the very first moments of Universe. The cosmological constant, however, has been introduced by Einstein for the sake of making Universe a steady one. The discovery that Universe expands ruined its credibility. On the other hand the inflationist model allows the cosmological constant to be given an upper bound of $\Lambda \leq 10^{-55}\ cm^{-2}$ which leads to a vacuum energy density, whose expression is:

$$\rho_{vac} = \frac{\Lambda c^4}{4\pi G_g},$$

of $\rho_{vac} \approx 10^{-6}\ ergs.cm^{-3}$. In unified theories such as the grand unification theory (GUT), however, this energy is $\rho_{vac} \approx 10^{112}\ ergs.cm^{-3}$. A discrepancy of 118 orders of magnitude is the sign that the nature of vacuum is misunderstood to say the least.

The metric scale could shed another light on this problem. Just as a neutron star may be considered as a huge nucleus (in the sense of nuclear physics) with the nuclear scale $(\cong 10^{-13}\ m)$ swelling up to the neutron stars scale $(\cong 10^4\ m)$, the early Universe should be a « cell » whose scale, either the metric scale $(\cong 10^{-24}\ m)$ or even the Planck scale $(\cong 10^{-34}\ m)$, blew up to some $(10\ m)$. In this early Universe all distances vanish so ensuring the homogeneity of the system.

**References:**

[1]-M.S. Schnubel, C.W. Akerlof, S. Biller et al. The Astrophysical Journal **460**, p.644 (1996)

[2]-J.R. Westlake, « A handbook of numerical matrix inversion and solution of linear equations » (John Wiley & Sons) pp. 126-127. (1968)

[3]-A discussion on the Landau-Ginzburg criterion may be found in the book of Mr. Le Bellac (and references therein), « Des phénomènes critiques aux champs de jauge: Une introduction aux méthodes et aux applications de la théorie quantique des champs » (Savoirs Actuels InterEdition/Editions du CNRS) pp. 75-82 (1990).

[4]-J. P. Serre, «Représentations linéaires des groupes finis » (Hermann, Paris) p. 165 (1978)

[5] M. Le Bellac, « Des phénomènes critiques aux champs de jauge: Une introduction aux méthodes et aux applications de la théorie quantique des champs » (Savoirs Actuels InterEdition/Editions du CNRS) pp. 71-73 (1990).

[6]-See for example:

J.R. Aitchison, « An informal introduction to gauge theory » Cambridge University Press (1982)

or J.C. Taylor, « Gauge theories of weak interactions » Cambrige University Press (1978)

or Mr. Le Bellac's book. pp.568-575.

[7] -J.W. Moffat: « Stochastic gravity » a preprint from the University of Toronto (Canada) UTPT-96-16 (1997)

[8]- A. Aspect, Phys.Rev.Lett. **49**, pp. 91 et 1804 (1982)

[9]- B.S. DeWitt and N. Graham, « The many-worlds interpretation of quantum mechanics » (Princeton University Press) (1973)